\documentclass{article}

\usepackage{graphicx}
\usepackage{hyperref}

\newcommand{\aj}{Astron. J.}

\newcommand{\aap}{Astron. Astrophys.}

\title{The decline of astronomical research in Venezuela}
\author{N\'estor S\'anchez \\
        Universidad Internacional de Valencia (VIU), Spain}

\date{\today}

\begin{document}

\maketitle

\section*{}

{\bf
During the last 15 years the number of astronomy-related
papers published by scientists in Venezuela has been 
continuously decreasing, mainly due to emigration. 
If rapid corrective actions are not implemented, 
Venezuelan astronomy could disappear.
}

\section*{}

Although its first steps date back to the XIX century, it can be
said that Astronomy in Venezuela started in the 1970s with the 
construction of the National Astronomical Observatory and the
foundation of the Centro de Investigaciones de Astronom\'{\i}a
(CIDA) created to manage the Observatory \cite{Mar07}.
During the 1980s, Venezuela was strengthening and consolidating
this research field until becoming a very competitive country in
Latin America, only behind the four countries with the longest
astronomical tradition in the region: Argentina, Brazil, Chile
and Mexico \cite{Hea07,Rib13}.

Since 1999 Venezuela started a transformation process, the
``Socialism of the 21st century", led by the State. Since
then, a considerable debate remains between supporters and
detractors of the Venezuelan government about whether the 
change has been for better or for worse, in a deeply polarised
country \cite{Mor15}. Science and Technology
have been part of this process and,
obviously, have also suffered from such a polarisation.
While some authors consider that Venezuelan government is
backing science and that, despite the political crisis,
the total investment has been continuously
increasing \cite{Cab03,Bar09,Cha09}, other researchers
claim that budget cuts, discretionary use of resources
(on the basis of political loyalty) and, in general, the
economic, political and social crisis are seriously damaging
scientific research in Venezuela \cite{Men03,Req03,Bif09}.
Apart from the existing polarization, the economic crisis
makes things difficult. Almost all research funding comes
from the State but most of the budget goes to pay salaries
and only a small percentage is left for infrastructures,
graduate or postgraduate programmes and research projects
\cite{Fra16}. Payments are made in local currency
(bol\'{\i}var) that suffers high inflation and
there is an exchange control system that prevents
unauthorized (by Venezuelan government) access to
foreign currency. All this makes it difficult to
researchers and institutions to buy needed resources
(materials, equipments, books and journals) or to
participate in international events and, additionally,
makes Venezuela unattractive as a destination for foreign
researcher or student mobility.

An apparently undeniable reality is that, whatever the
underlying reasons, Venezuelan scientists (by this I mean
scientists based in Venezuela and working for a Venezuelan
institution, irrespective of their nationality) are emigrating
at accelerated rates \cite{Req16a,Req16b} and, as a consequence,
scientific production (published papers) has been decreasing 
over the last years \cite{Pan14}. The brain drain seems to be 
similar in magnitude for all fields of knowledge \cite{Req16b} 
and for Venezuelan Astronomy in particular has meant a setback 
of about $20$ years.

\section*{40 years of Astronomy in Venezuela}

According to the NASA Astrophysics Data System (ADS) \cite{Kur00},
between the years 1980 and 2019 inclusive, a total of $798$
refereed papers in the astronomy collection (excluding 
proceedings and book chapters) were published with at
least one author affiliated to a Venezuelan institution.
Table~\ref{tabpapers} summarizes the main characteristics
derived from these publications.

\begin{table}[ht] 
\centering
\begin{tabular}{| l r |} 
\hline
\multicolumn{2}{|c|}{\bf Main research networks} \\
\hline
Stellar, Galactic and Extragalactic Astronomy & 37.3\% \\
Relativistic Astrophysics (and Theoretical Physics) & 31.1\% \\
Atomic Physics & 7.4\% \\
Fluid mechanics & 6.3\% \\
\hline
\multicolumn{2}{|c|}{\bf Main journals} \\
\hline
The Astrophysical Journal & 25.2\% \\
Physical Review D & 10.3\% \\
Astronomy and Astrophysics & 8.8\% \\
Monthly Notices of the Royal Astronomical Society & 8.0\% \\
Astrophysics and Space Science & 5.9\% \\
Classical and Quantum Gravity & 5.5\% \\
General Relativity and Gravitation & 4.8\% \\
The Astronomical Journal & 4.1\% \\
\hline
\multicolumn{2}{|c|}{\bf Main research centres} \\
\hline
Centro de Investigaciones de Astronom\'{\i}a (CIDA) & 29.9\% \\
Universidad de Los Andes (ULA) & 22.2\% \\
Universidad Central de Venezuela (UCV) & 16.3\% \\
Universidad Sim\'{o}n Bol\'{\i}var (USB) & 13.7\% \\
Instituvo Venezolano de Investigaciones Cient\'{\i}ficas (IVIC) & 11.2\% \\
\hline
\end{tabular}
\caption{Main characteristics of Astronomy/Astrophysics
research in Venezuela (years 1980-2019).}
\label{tabpapers}
\end{table}

By checking out these publications and using other ADS tools,
it is possible to identify four main (not the only ones)
research networks. i.e., group of authors sharing papers and/or
research lines. The most important network is related to what we
could call ``traditional" Astronomy, with a bulk of papers
dealing with Stellar Astronomy (including young stars
and variable stars), Galactic Astronomy (including stellar
clusters) and Extragalactic Astronomy. This research was
published in standard astronomical journals (especially
The Astrophysical Journal) and it was carried out at
CIDA (mostly) and ULA.
There is a second important but completely different
network whose major research line is Relativistic 
Astrophysics (relativistic fluids), although this 
network also includes several papers and authors 
working in different areas of Theoretical Physics 
and publishing, logically, in different journals,
especially Physical Review D. Authors in this network
belong to the following institutions (in order of
contributions): UCV, ULA, USB and IVIC.
These two main networks account for almost $70$\%
of the total Astronomy research in Venezuela in
the last 40 years. There are, however, two smaller
but very well-defined research networks: one refers
to Atomic Physics (atomic data for astrophysics)
with researchers mainly working at IVIC, and the
other refers to Fluid Mechanics (applied to
astrophysical plasmas) with researchers mainly
at ULA. There are other institutions, in particular
universities, doing some research in Astronomy, but
their total contributions (refereed papers) are always
below $\sim 2.5$\%.

\begin{figure}[t]
\includegraphics[width=\textwidth]{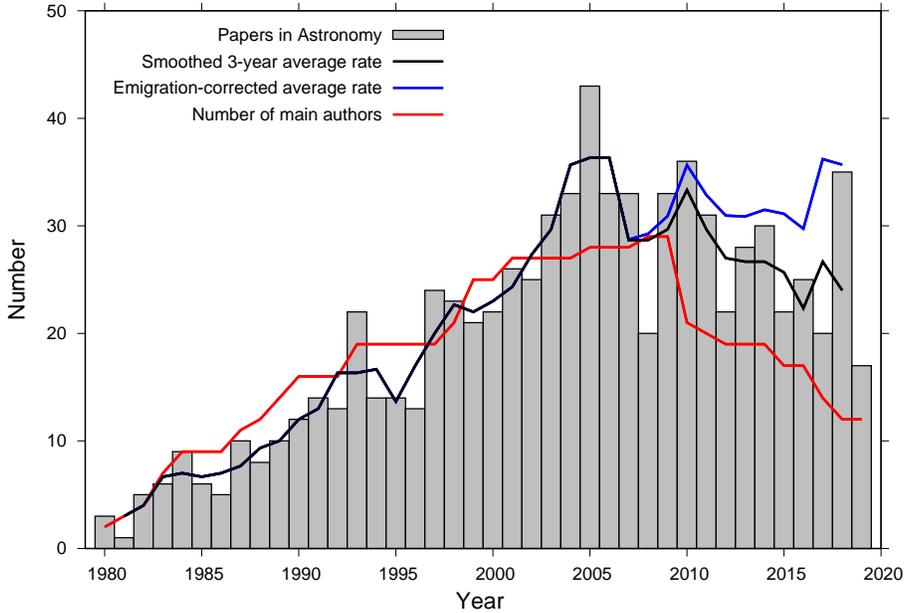}
\caption{Evolution of Astronomy research in Venezuela.
The histogram is the number of refereed papers with at
least one author affiliated to a Venezuelan institution
and the black line is a smoothed version
(3-year bin-averaged data) that better shows global
trends. Red line indicates the total number of main authors
(see text) that were active per year, which started to
decrease in 2009 due to emigration of astronomers.
Blue line is the estimated production if there had been
no emigration flow.}
\label{figpapers}
\centering
\end{figure}

The number of Astronomy-related papers per year was steadily
increasing from 1980 to 2005 (see Figure~\ref{figpapers}).
The publication rate in the peak (on average, between 2004
and 2006) was about $35$ papers/year. From there, beyond
yearly variations, there is a clear decreasing trend that
continues until today. The mean publication rate between
2018 and 2019 was $\sim 25$ papers/year, a $\sim 30$\%
decline from the peak in 2005. Current production rate
is equivalent to mean production around year 2000
($\sim 20$ years ago). Interestingly, publications
in Astronomy started decreasing several years before
total scientific publications in Venezuela that,
according to Science Citation Index and Scopus,
started decreasing in 2008 \cite{Pan14}.

\section*{Venezuelan astronomers are emigrating}

By using the full list of papers (and ADS tools), I searched
for the main Venezuelan astronomers/astrophysicists.
By ``main" I mean authors who published at least 10 papers
betweem 1980 and 2019 or, if not fulfilling this condition,
who have an average publication rate $\ge 0.5$ papers/year.
I extracted the list of main authors and also their
institutions and their activity time intervals (years
of the first and last papers published with affiliation
to a Venezuelan institution). I identified $30$ main
authors, from which $73$\% worked at CIDA (14 authos)
or ULA (8) and the remaining were affiliated with other
centres (IVIC, UCV, USB). This group of main authors
have been involved in $\sim 65$\% of the papers
published during the last $40$ years. The number
of years of author's activity ranges from 8 to 40
(still working in Venezuela) following a roughly
flat distribution. Their mean publications rates
ranges from 0.5 to 3.5 papers/year with $47$\% of
the authors publishing more than $1$ paper/year and
only $13$\% more than $2$ papers/year (on average).

Once identified these main authors, we use the ADS
to verify whether, after their last active year,
they have or have not continued publishing with 
affiliation to a different country. Since 2007,
$12$ of the main authors have emigrated to other
countries (Chile, Colombia, Mexico, Spain and USA),
$5$ have stopped publishing at all (retirement or
unknown reasons) and $2$ have passed away. There 
are still $11$ active researchers in Venezuela,
though I know for sure that at least $4$ of them
are actually living abroad but they still sign
papers with affiliation country Venezuela.
The number of active authors per year is plotted
as red line in Figure~\ref{figpapers}.
The emigration process started in 2007 but the
number of new main authors increased until 2008,
so the net number of active researchers started
decreasing since 2009-2010. The number of main
authors in 2019 is only a $38$\% of the historical
peak value in 2009.
Remarkable is the fact that there are no new main
astronomers in Venezuela since 2008, based on the
mentioned criteria of having an average production
rate higher than 0.5 Astronomy-related refereed
papers per year.

What has been the real impact of the emigration
of Venezuelan astronomers? 
By knowing the mean production rates and emigraton
years for each of the $12$ emigrated authors, the
total mean production rate can be corrected by
emigration (blue line in Figure~\ref{figpapers}).
Despite the yearly variations, the corrected number
of papers per year remains nearly constant over the
last $10$ years around $33.1$ papers/year, a value
very close to the historical maximum.

\section*{Venezuelan Astronomy is falling behind}

\begin{figure}[t]
\includegraphics[width=\textwidth]{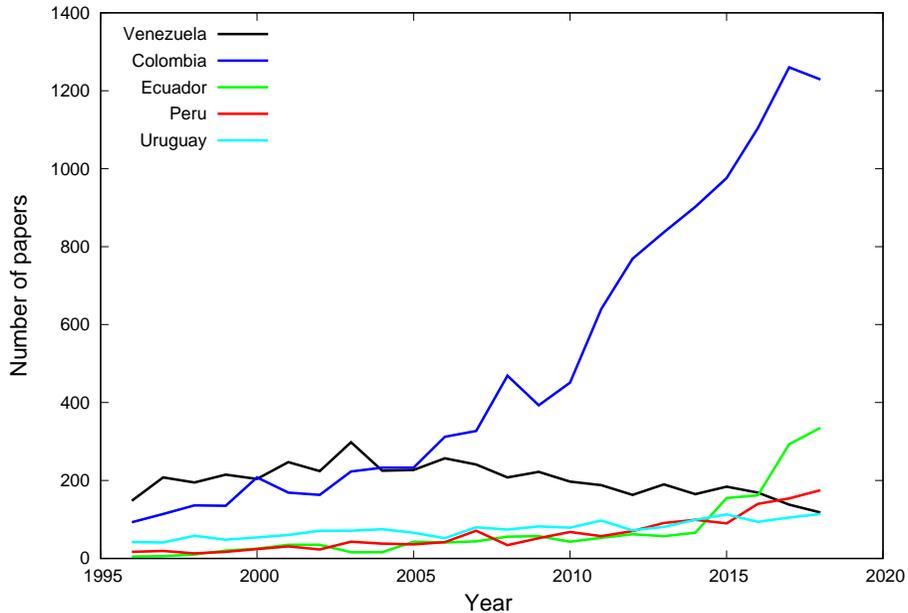}
\caption{Comparison of publications in the area of 
Physics and Astronomy in Venezuela with those in
other countries in Latin America according to
SCImago Journal \& Country Rank.}
\label{figcountries}
\centering
\end{figure}

A direct comparison with other Latin American countries 
clearly illustrates the gradual destruction of scientific
research in Venezuela.
Figure~\ref{figcountries} shows the number of papers
in Physics and Astronomy for some selected countries
(excluding the top 4 countries in Astronomy research
in Latin America). The data was retrieved from 
\href{http://www.scimagojr.com}{SCImago Journal \& 
Country Rank}.
Venezuela is the only country that has inverted the
general growing trend and Colombia, Ecuador, Peru
and Uruguay have already surpassed Venezuela in
number of papers per year.
Particularly notorious is the case of Colombian
Astronomy \cite{Hig17}, for which a clear slope
change can be seen that it has positioned in the
fifth place in the region since 2005.

In terms of production rate (papers per year), Venezuelan
Astronomy is at present equivalent to that $\sim 20$ years
ago. The number of papers depends almost exclusively on the
number of researchers \cite{Abt06}. So, the true problem of
Venezuelan Astronomy is not the current production rate
but the fact that Venezuelan scientists are emigrating
and this emigration is far from being stopped. According
to data from \href{https://www.worldbank.org}{World Bank}
the total population in Venezuela started to decrease for
the first time in its history, from an historical maximum
of $\sim 30$ millions of inhabitants in 2015 to $\sim 28.9$
millions in 2018. And the main reason for this drop is the
net migration flow because, even though the birth rate has
been decreasing for several decades, it is still higher
than the (currently increasing) death rate. The
\href{https://www.iom.int}{International Organization
for Migration} estimates that in 2017 the number of 
Venezuelans living abroad was $1.6$ millions.
Astronomy is a particularly fragile community in Venezuela
because of the relatively low number of senior researchers.
A simple linear extrapolation (by eye) suggests that
Venezuelan Astronomy could even disappear entirely in
a few more decades.

Can this trend be reversed? Certainly not in the short term.
Venezuelans scientists are not leaving their country because
of difficulties in their scientific activities, but rather because of
serious difficulties in their day-to-day lives. Their salaries seem
ridiculous in a country with a hyperinflation that, according to
official estimates of the Central Bank of Venezuela, decreased
from $\sim 130.000$\% in 2018 to ``only" $\sim 9.600$\% in 2019.
In any case, money is pretty useless due to a serious scarcity of
food, medicines and other basic goods. Violence is also a
contributing factor, being Venezuela one of the most dangerous
countries in the world. The non-governmental organization known
as the \href{https://observatoriodeviolencia.org.ve}{Venezuelan
Violence Observatory} estimated that in 2019 about $16.500$
people were killed (homicides committed by criminals, caused
by ``resistance to authority" or violent deaths of undetermined
intent), placing Venezuela as the country with the highest per
capita murder rate in Latin America.
Under this scenario, to define new science policies and/or to
notoriously increase the assigned budget in order to improve 
infrastructures or graduate and research programmes is not 
going to revert the situation.
Deeper changes must be undertaken to address the underlying
causes. Venezuela's political, economic, and social crisis has
to be overcome in the first place. After that, it will probably
take a long time to recover what Venezuela once had.

\end{document}